\documentclass[final,5p,times,twocolumn]{elsarticle}

\usepackage{amssymb}
\usepackage{amsmath}
\usepackage{algorithm}
\usepackage{algorithmic}
\usepackage{caption}
\usepackage{subcaption}
\usepackage{threeparttable}
\usepackage{multirow}
\usepackage{color}
\usepackage[color]{changebar}
\cbcolor{red}

\DeclareMathAlphabet{\mathpzc}{OT1}{pzc}{m}{it}
\newcommand{\pe}{$\mathpzc{pe}$}

\journal{Computer Physics Communications}

\begin{document}

\begin{frontmatter}



\title{Simulation of a Hard-Spherocylinder Liquid Crystal with the \pe}


\author[ThPhy]{Ellen Fischermeier\corref{cor1}}
\ead{ellen.fischermeier@fau.de}
\cortext[cor1]{Corresponding author. Tel.: +49 91318528450}
\author[LSS]{Dominik Bartuschat}
\author[LSS]{Tobias Preclik}
\author[ThPhy]{Matthieu Marechal}
\author[ThPhy]{Klaus Mecke}

\address[ThPhy]{Institut f\"{u}r Theoretische Physik I, Friedrich-Alexander Universit\"{a}t Erlangen-N\"{u}rnberg, Staudtstra{\ss}e 7, D-91058 Erlangen, Germany}
\address[LSS]{Lehrstuhl f\"{u}r Systemsimulation, Friedrich-Alexander Universit\"{a}t Erlangen-N\"{u}rnberg, Cauerstra{\ss}e 11, D-91058 Erlangen, Germany}

\begin{abstract}
The \pe{} physics engine is validated through the simulation of a liquid crystal model system consisting of hard spherocylinders.
For this purpose we evaluate several characteristic parameters of this system, namely the nematic order parameter, the pressure, 
and the Frank elastic constants. We compare these to the values reported in literature and find a very good agreement, which demonstrates that the \pe{} physics engine can accurately treat such densely packed particle systems.
Simultaneously we are able to examine the influence of finite size effects, especially on the evaluation of the Frank elastic constants,
as we are far less restricted in system size than earlier simulations.
\end{abstract}

\begin{keyword}
Parallel framework \sep Rigid body dynamics \sep Hard rod \sep Liquid crystal \sep Frank elastic constants  \PACS 07.05.Tp \sep 61.30.-v


\end{keyword}

\end{frontmatter}


NOTICE: This is the author's version of a work that was accepted for publication in Computer Physics Communications.
Changes resulting from the publishing process, such as peer review, editing, corrections, structural formatting,
and other quality control mechanisms may not be reflected in this document.
Changes may have been made to this work since it was submitted for publication. 
A definitive version was subsequently published in Computer Physics Communications, [VOL 185, ISSUE 12, (2014)] DOI 10.1016/j.cpc.2014.08.014
\section{Introduction}
\label{sec:introduction}

The \pe{} physics engine~\cite{Iglberger_Rigid:2009,Iglberger_Software:2010} is a software framework for simulating large-scale multi-body systems in soft and hard contact. The simulated
bodies have spatial extension and their core shapes are assumed to be rigid. The software supports various numerical methods to resolve
the multi-contact problems such as discrete element methods (DEM)~\cite{Cundall_dicsrete:1979} for soft
contacts and the fast frictional dynamics (FFD)~\cite{Kaufman_Fast_article:2005} and non-smooth contact
dynamics (NSCD)~\cite{Jean:1999} methods for hard contacts.
The physics engine was successfully used to simulate large-scale granular flows~\cite{Iglberger_Massively:2010} on its own but also
coupled to the \textsc{waLBerla} lattice Boltzmann framework~\cite{Feichtinger_walberla:2011} for the simulation of particulate
flows~\cite{Goetz_Direct:2010} and self propelled swimmers~\cite{Pickl_All:2011}. The software framework was designed from the ground
up targeting distributed-memory architectures and has been shown to scale well up to the full size of today's largest
supercomputers~\cite{Iglberger_Massively:2010,Preclik_Ultrascale:2014,Pickl_All:2011}. In particular, the FFD method qualifies well for the parallelization since the contact problems are treated in a localized
manner keeping the communication amount to a minimum. The FFD method is originally based on Moreau's midpoint rule, an established time-stepping
scheme for simulating mechanical systems in frictional contact~\cite{Moreau_Panagiotopoulos:1988}.
It uses a time-integrator equivalent to a Leapfrog Verlet integration if no contacts are present.
The friction model is derived from the principle of maximal dissipation and captures essential frictional properties~\cite{Kaufman_Fast_article:2005}.
However, the contact problem simplification stemming from the localization and the special type of friction model require a
thorough validation of the numerical method. This paper performs such a validation on the basis of a hard-spherocylinder system as a model
for liquid crystals, where the implementation of the FFD method within the \pe{} framework replaces conventional molecular dynamics techniques.

Studies of hard-spherocylinder systems date back to the '70s and the work of Few and Rigby~\cite{Few_Equation:1973} and Vieillard-Baron~\cite{Vieillard-Baron_equation:1974}.
Spherocylinders consist of a cylinder of length $L$ and diameter $D$ with hemispherical end caps. For $L/D=0$ this model system corresponds to the hard-sphere model whereas 
for $L/D\rightarrow\infty$ the system consists of infinitely thin needles. Systems of rigid straight rods have also been realized experimentally for example in the form of 
tobacco mosaic viruses~\cite{Bernal_xray:1941} and colloidal silica rods~\cite{Kuijk_phase:2012}.

\begin{figure}[h!]
        \centering
        \includegraphics[width=8.86cm]{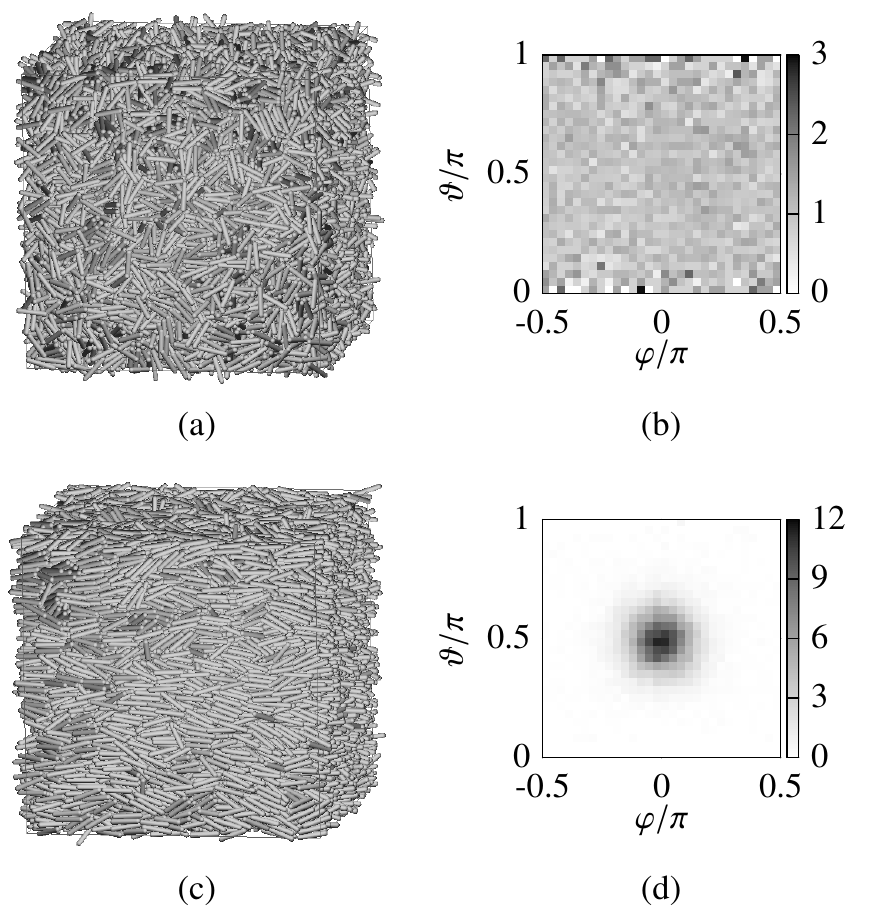}
        \caption{Snapshots of the $L/D=5$ hard-spherocylinder system in (a) an isotropic ($\rho^*=0.4$) and (c) a nematic ($\rho^*=0.5$) state 
        with the corresponding angular distribution profiles ((b) isotropic, (d) nematic) normalized to the ideal isotropic distribution profile. (For definition of dimensionless density $\rho^*$ see Section~\ref{sec:setup}.)
        The increased noise around $\vartheta=0$ and $\vartheta=\pi$ in (b) is due to the reduced statistics in this regime as the size of the solid 
        angle bins behaves like $\sin(\vartheta)\mathrm{d}\vartheta \mathrm{d}\varphi$.}
    \label{fig:snapshots}
\end{figure}

All phase transitions of such a system are purely driven by entropy as the hard particles interact exclusively via excluded volume. Therefore, the temperature $k_B T$ acts only as an energy scale. 
Two types of entropy compete here: The orientational entropy drives the system towards an isotropic phase where both center of masses and orientations are uniformly distributed 
(see Fig.~\ref{fig:snapshots}a,\,b). The second type of entropy, translational entropy, favors a uniform orientation of the particles which minimizes the excluded volume. An example of such an ordered phase is the nematic phase which is characterized by alignment 
of the particles with respect to each other, giving rise to an anisotropic angular distribution profile while keeping a homogeneous distribution of the center of masses (see Fig.~\ref{fig:snapshots}c,\,d). This combination of liquid-like translational and crystal-like orientational properties 
which extends to many observable physical properties is also the origin of the name liquid crystal.  

For $L/D\rightarrow\infty$ in the Onsager limit~\cite{Onsager_effects:1949} the hard-spherocylinder system exhibits a phase transition from isotropic to 
nematic at vanishing volume packing fraction (order $D/L$). 
For systems with a finite aspect ratio $L/D$ and full translational and rotational freedom Frenkel and co-workers were the first to report such a phase 
transition with molecular dynamics~(MD) and Monte Carlo~(MC) simulations~\cite{Frenkel_Structure:1988,Frenkel_Thermodynamic:1988}. They also discovered a crystalline solid phase and 
a smectic A phase where the system exhibits layering of the particles, thus having a one-dimensional positional ordering in addition to the orientational ordering.
In their later work~\cite{Veerman_Phase:1990,Bolhuis_Tracing:1997} they demonstrated that the hard-spherocylinder system indeed exhibits a rich phase diagram which depends 
on the aspect ratio of the particles as well as on the packing fraction, as confirmed by the studies of McGrother et al.~\cite{McGrother_re‐examination:1996}.

This well understood system provides an ideal test case for our simulation framework, the \pe{} physics engine, in the context of micro-scale 
multi-particle systems.
The massive parallelism of the \pe{} software framework
enables us to study far larger systems than those considered in the publications mentioned above. Those were restricted to a few hundred particles with only a couple of exceptions 
which included up to a few thousand particles. We can easily simulate systems of roughly $20$ times that number of particles without straining the limits of computational power. This gives us
the possibility to improve the confidence level of the results for the observables studied in such a hard-spherocylinder system. We focus on the the $L/D=5$ system at a single packing fraction in the 
nematic phase where the particles are rather densely packed. As it is known that equilibration and numerical treatment become difficult in this regime, these are ideal conditions for testing
the power of our framework.

The paper is structured as follows: Section~\ref{sec:numerical} describes the \pe{} rigid body physics engine in detail and depicts both,
the simulation setup, and the specifics of the model system we employ. In Section~\ref{sec:theory}, we elaborate on the variables we evaluate in this work, 
which are the order parameter, the pressure, and the Frank elastic constants. We further give a short overview over their theoretical description.
The results of this work are presented in Section~\ref{sec:results} and some conclusions are drawn in Section~\ref{sec:conc}.

\section{Numerical methods and simulation setup}
\label{sec:numerical}
\subsection{The rigid body physics engine \pe{}}
The \pe{} physics engine is a powerful tool for large-scale rigid body dynamics simulations~\cite{Iglberger_Rigid:2009,Iglberger_Software:2010}.
The algorithm we chose to employ in this work is based on the FFD solver first proposed by Kaufman et al.~\cite{Kaufman_Fast_article:2005,Kaufman_Fast_proceedings:2005},
where we set all friction coefficients to zero to make our results comparable to previous studies. It is parallelized with MPI~\cite{Gropp_Using:1999} following a domain decomposition approach: Each process is responsible for the time-integration of the bodies whose center of mass is located in its associated computational
subdomain. In addition to these local bodies processes obtain shadow copies of bodies intersecting their subdomain. Algorithm \ref{algo:pe} lists pseudo-code for
the time step procedure.

\begin{algorithm}[h!]
   \caption{Rigid Body Time Step of Size $dt$~\cite{Iglberger_Rigid:2009,Iglberger_Software:2010}}
   \label{algo:pe}
   \begin{algorithmic}[1]
      \STATE{// 1.~MPI message exchange: Reduce forces and torques}
      \STATE{}
      \STATE{// First half time step}
      \FOR{\textit{each local rigid body ${B_i}$}}
        \STATE{\textit{advance position and orientation for $\frac{dt}{2}$}}
        \STATE{\textit{advance translational and angular velocity for $\frac{dt}{2}$}}
      \ENDFOR
      \STATE{}
      \STATE{// 2.~MPI message exchange: Synchronize rigid bodies}
      \STATE{}
      \STATE{// Collision detection}
      \FOR{\textit{each local rigid body and shadow copy ${B_i}$}}
        \STATE{\textit{detect all contacts $C(B_i)$}}
        \FOR{\textit{all contacts $k\in C(B_i)$}}
        \STATE{\textit{determine acting constraints}}
        \ENDFOR
      \ENDFOR
      \STATE{}
      \STATE{// 3.~MPI message exchange: Gather contact constraints}
      \STATE{}
      \STATE{// Collision resolution and second half time step}
      \FOR{\textit{each local rigid body ${B_i}$}}
        \IF{${B_i}$ has constraints}
        \STATE{\textit{find post-collision translational and angular velocity}}
        \ELSE
        \STATE{\textit{advance translational and angular velocity for $\frac{dt}{2}$}}
        \ENDIF
        \STATE{\textit{advance position and orientation for $\frac{dt}{2}$}}
      \ENDFOR
      \STATE{}
      \STATE{// 4.~MPI message exchange: Synchronize rigid bodies}
   \end{algorithmic}
\end{algorithm}

The time step starts by sending all forces and torques applied to shadow copies to the respective owner process of the body. The
owner process can then reduce them and proceed by performing the time-integration up to the midpoint of the time step neglecting contact constraints
on purpose. Subsequently, the shadow copies need to be updated to the new positions and velocities and migrations of bodies to neighboring processes have to be
treated in a second message exchange communication step. Next, collisions are detected and filtered. Some of the collisions can only be detected by
processes holding shadow copies. These contact constraints are gathered in the third message exchange step at the processes responsible for the time-integration
of the bodies enabling them to compute the post-collision velocities prescribed by the FFD's contact model \cite{Kaufman_Fast_article:2005} and performing the second half of the time-integration.
Finally, the positions and velocities have to be synchronized again in the fourth and last message exchange communication step.

Particle overlaps are inherent in this algorithm and indeed collisions are only recognized if the particles intersect. 
Splitting the time step into two half steps, one before and one after the collision resolution, ensures that after the full time step the detected overlaps have been eliminated.
Still, the size of the time steps has to be chosen carefully (see Sec.~\ref{sec:setup}) to avoid a strong deviation from the expected hard-particle behavior 
due to the seemingly reduced volume packing fraction.    
The advantage of a fixed time step lies in the fact that time integration can be driven forward efficiently despite high collision frequencies in contrast to event-driven methods~\cite{McNamara_Inelastic:1994}.
Furthermore the code can be very efficiently parallelized as collisions are always treated locally.

\subsection{Simulation setup}
\label{sec:setup}
The model system used in this work consists of $N=18432$ hard spherocylinders of unity mass with an aspect ratio $L/D=5$. The unit of energy was set to $k_BT$ 
and the unit of length to $D$. The particles were enclosed in a cubical box of side length $l=57.0512\,D$ with periodic boundary conditions. This results in a dimensionless density of $\rho^*=0.5$ 
where the system is expected to be in the nematic phase~\cite{Frenkel_Structure:1988,Frenkel_Thermodynamic:1988,Veerman_Phase:1990,Bolhuis_Tracing:1997,McGrother_re‐examination:1996} 
with $\rho^*=\rho/\rho_{cp}$, $\rho_{cp}$ being the density in the close packing limit $\rho_{cp}=(2 D^{-3})/(\sqrt{2}+\frac L D \sqrt{3})$.

In the initial configuration the spherocylinders were arranged on a $8\times48\times48$ (tetragonal) lattice and had a uniform orientation along the x-axis.
Their initial velocity and angular velocity distributions were chosen randomly from Gaussian distributions such that the condition $k_BT=1$ held in our units. 
For reason of symmetry no angular momentum was applied around the symmetry axis of the particles, so the angular velocity is always perpendicular to this axis.
The time step was set to $dt=2\times10^{-6}\,\tau$ where the unit of time follows directly from our choice of unit energy, unit length and unit mass: 
$\tau=D\sqrt{m_{\text{SC}}/k_BT}$, with $m_{\text{SC}}$ the mass of a spherocylinder.

To provide an upper bound on the amount of overlap between particles in a collision, we consider the worst-case scenario in which two particles 
already touch in a point $\vec{p}$ at the beginning of the time step and define overlap as the surface to surface distance measured along the surface normal at $\vec{p}$ after the first half step.
If we consider only translational motion, this overlap is less than $10^{-5} D$ for all but a tiny fraction ($< 2\times10^{-12}$) of all collisions at this step size. The contribution due to rotation 
has the same order of magnitude. This clearly justifies the assumption of a hard-particle system, considering that this scenario poses the extreme case of collision.

A simple thermostat was used to correct for accumulation of numerical errors in the resolution of collisions.
The total kinetic energy was rescaled to its initial value every 200 time steps by rescaling angular velocity and velocity of each particle by the same fraction.
This rate of rescaling events suffices to suppress fluctuations of the kinetic energy and thereby keeping the temperature constant (the potential energy being always zero in the hard-particle system). 

\begin{figure}[h!]
 \centering
 \includegraphics[width=7.53cm]{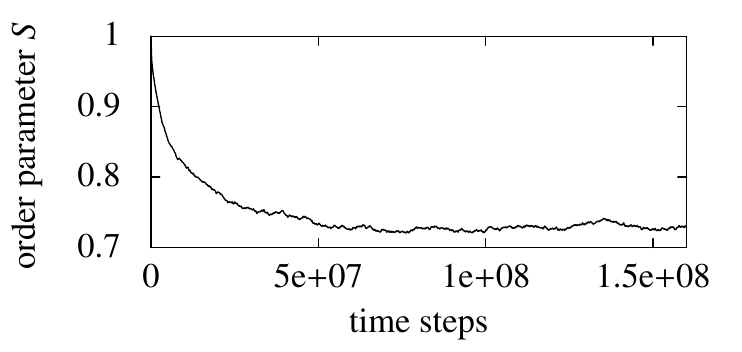}
  \caption{Relaxation of the system from perfect orientational order ($S=1$). After $10^8$ time steps the system is assumed to have reached equilibrium. 
 (For definition of the order parameter~$S$ see Eq.~\ref{eqn:Orderparam1}.)}
\label{fig:equilib} 
\end{figure}

To obtain uncorrelated results we performed six simulations with independently generated initial velocity configurations. The systems were allowed to relax for about $10^8$ time steps 
after which the nematic order parameter (see Sec.~\ref{subsec:orderparameter}) indicated the systems to be in equilibrium. 
A typical equilibration curve is shown in Fig.~\ref{fig:equilib}, a snapshot of the system in the nematic phase 
is presented in Fig.~\ref{fig:snapshots}c).
Subsequently, production runs of about $5\times10^7$ time steps were performed the results of which are presented in Sec.~\ref{sec:results}.

\section{Physical quantities of interest}
\label{sec:theory}

\subsection{Order parameter}
\label{subsec:orderparameter}
The first parameter we use to characterize the system is the order parameter $S$. 
It is a measure for the degree of nematic ordering in the system, i.e., the alignment of the particles along a common direction, the nematic director $\vec{n}$.
Microscopically it is defined by the average of the second Legendre polynomial of the cosine of the angle $\theta$ between 
the director $\vec{n}$ and the symmetry axis of the individual spherocylinders $\vec{e}_i$:
\begin{equation}
 S=\left<P_2\left(\cos\theta\right)\right>=\frac{3}{2}\int_{0}^{\pi}\cos^2\theta g(\theta)\sin\theta\mathrm{d}\theta-\frac{1}{2}
\label{eqn:Orderparam1}
\end{equation}
with the equilibrium orientational distribution function $g(\theta)$. $\left<...\right>$ indicates both the equilibrium ensemble average and the average over all particles in the system.

In the simulation however, the nematic director is not known a priori,
thus the angular distribution with respect to this vector can't be measured directly.
Instead of $g(\theta)$ we consider therefore the order tensor $\boldsymbol{Q}$:
\begin{equation}
Q_{\alpha\beta}=\frac 1 N \left(\sum_i \frac 3 2 e_{i\alpha} e_{i\beta}\right)-\frac 1 2 \delta_{\alpha\beta} \,\, , \,\,\, \,\,\,\alpha,\beta=x,y,z\;.
\label{eqn:Orderparam}
\end{equation}
The sum goes over all particles in the system, $N$ being their total number, with $\delta_{\alpha\beta}$ the Kronecker delta. 
This tensor can be calculated directly from the single particle orientations $\vec{e}_i$ and it can be shown that its largest eigenvalue corresponds to the order parameter $S$ 
as defined above~\cite{Gennes:1974,Eppenga:1984}. Furthermore the associated eigenvector yields the nematic director $\vec{n}$.
The order parameter can take values between $0$ and $1$, a value of $S=0$ indicating an isotropic distribution of particle orientations 
whereas $S=1$ is reached only in the case of perfect alignment of all particles towards each other and thus towards the director.

\subsection{Pressure}
\label{subsec:pressure}
\begin{figure}[h]
 \centering
 \includegraphics[width=5cm]{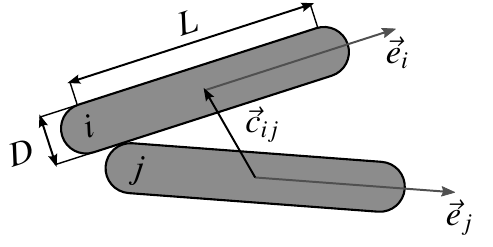}
 \caption{Sketch of two colliding spherocylinders}
 \label{fig:particlePos}
\end{figure}

A second property of the system we evaluate is the pressure. According to the virial equation it can be calculated as
\begin{equation}
P=\frac 1 \beta \frac{N}{V}+\frac{1} {3V}\left<\sum_{i<j}^N \vec{F}_{ij} \vec{c}_{ij}\right> 
\end{equation}
with the usual definition $\beta^{-1}=k_B T$, $N$ the total number of particles, $V$ the considered volume and $\vec{F}_{ij}$ the force exerted by particle $j$ on particle $i$, 
$\left<...\right>$ indicating the equilibrium ensemble average~\cite{Frenkel_Understanding:1996}. The vector $\vec{c}_{ij}$ connects the center of mass positions of 
the two particles, i.e., $\vec{c}_{ij}=\vec{r}_i-\vec{r}_j$ (see Fig.~\ref{fig:particlePos}).

For hard bodies the forces $\vec{F}_{ij}$ are not continuous as the inter-particle potential is either $0$ in case of no overlap or $\infty$ if the particles intersect.
Here the forces are substituted by the transferred momentum per unit time~\cite{Rebertus_Molecular:1977}: 
\begin{equation}
  P=\frac 1 \beta \frac{N}{V}+\frac{1} {3V\tau}\sum_{C}\Delta\vec{p}_{ij} \vec{c}_{ij}\;. 
\label{eqn:pressure}
\end{equation}
The sum is taken over all collisions $C$ within the observation time $\tau$ and $\Delta\vec{p}_{ij}$ is calculated from the difference between the velocity of particle $i$ before and after 
resolution of the contact with particle $j$.

\subsection{Frank elastic constants}
\label{subsec:frank}
The Frank elastic constants $K_i$ ($i=1,2,3$) describe the free energy penalty for elastic deformation. There are three types of deformation to be considered,
namely, splay, twist and bend. For small deformations this additional free energy can be expressed in terms of the derivatives of the director field $\vec{n}(\vec{r})$ 
which represents the local orientation of the nematic director~\cite{Gennes:1974,Frank_Liquid:1958}:
\begin{multline}
     \Delta F \left[\vec{n}(\vec{r})\right]=\\ \frac{1}{2}\int \mathrm{d}\vec{r}\left(K_1(\nabla \cdot \vec{n})^2+K_2(\vec{n}\cdot(\nabla \times \vec{n}))^2+K_3(\vec{n}\times(\nabla\times\vec{n}))^2\right)
\end{multline}
The Frank constants are the proportionality constants for the contribution of the different types of deformation. $K_1$ is attributed to splay, $K_2$ to twist and $K_3$ to bend.

We employ a well established method to obtain the Frank constants from simulation which utilizes the Fourier transform of the order tensor $\boldsymbol{Q}$ defined 
in Section~\ref{subsec:orderparameter}~\cite{Allen_Calculation:1988,Allen_Erratum:1990,TjiptoMargo_Elastic:1992,Allen_Molecular:1996,OBrien_Elastic:2008}.
Those Fourier components $\tilde{Q}_{\alpha\beta}(\vec{k})$
\begin{equation}
 \tilde{Q}_{\alpha\beta}(\vec{k})=\frac V N \sum_i \left(\frac 3 2 e_{i\alpha} e_{i\beta}-\frac 1 2 \delta_{\alpha\beta}\right)\exp(\text{i}\vec{k}\vec{r}_i)\,\, , \,\,\, \,\,\,\alpha,\beta=1,2,3
\label{eqn:FourierOrdertens}
\end{equation}
are measured in a basis in which the 3-direction is given by the nematic director and $\vec{k}$ lies in the 1-3-plane.
In this coordinate system, the components of the Fourier transformed order tensor are at low $k$ related to the Frank elastic constants via
\begin{equation}
    E_{\mu3}=\frac{\frac{9}{4}S^2 V k_B T}{\left<\left|\tilde{Q}_{\mu 3}(\vec{k})\right|^2\right>}\rightarrow K_\mu k_1^2+K_3 k_3^2  \text{\qquad for \qquad} k\rightarrow 0
\label{eqn:Emu}
\end{equation}  
with $\mu=1,2$, $\left<...\right>$ again indicating the equilibrium ensemble average~\cite{Forster_Hydrodynamic:1975}.

\section{Results}
\label{sec:results}
\subsection{Evaluation of the nematic order parameter}
\label{subsec:resOrderparam}
The order parameter values were obtained every $10000$ time steps by calculating the order tensor $\boldsymbol{Q}$ (see Eq.~\ref{eqn:Orderparam}) and obtaining its highest eigenvalue $S$. 
The corresponding eigenvector, the nematic director $\vec{n}$, was also evaluated as it is required for the calculation of the Frank elastic constants (see Section~\ref{subsec:resFrank}).
A higher sampling frequency would not have contributed further information as subsequent values in a molecular dynamics simulation are highly correlated. 

The nematic order parameter was obtained independently for the different initial configurations by averaging over about $5000$ values each.
Averaging again over these uncorrelated values results in an average order parameter of $0.727\pm0.002$. This is in very good agreement with results from earlier MD and 
MC simulations~\cite{Bolhuis_Tracing:1997,McGrother_re‐examination:1996,Allen_Calculation:1988} as well as from density functional theory studies (DFT)~\cite{Poniewierski_Density:1990,Sin-Doo_Density:1989} (see Tab.~\ref{tab:Frank}).
\subsection{Evaluation of pressure}
\label{subsec:resPressure}
In this section we present our results on the pressure in terms of the rescaled pressure $P^*=\beta v_0 P$,
 where $v_0=\pi D^2\left(D/6+L/4\right)$ is the molecular volume of a spherocylinder. For each of our individual initial setups we calculated the pressure by evaluating $\sum \Delta\vec{p}_{ij} \vec{c}_{ij}$ 
over the collisions within a single time step $dt$ every $1000$ time steps. These results were then added up, making the observation time $\tau$ equal to $dt$ times the number of data points taken.
For each of the initial configurations the pressure was calculated separately according to Eq.~\ref{eqn:pressure}.
We then took the average over the pressure values from the independent simulations, obtaining an average rescaled pressure of $6.015\pm0.020$ which agrees very well with the 
values from~\cite{Frenkel_Structure:1988,Bolhuis_Tracing:1997,McGrother_re‐examination:1996} (see Tab.~\ref{tab:Frank}).
\subsection{Evaluation of Frank elastic constants}
\label{subsec:resFrank}
\begin{figure}[h!]
        \centering
           \includegraphics[width=8.73cm]{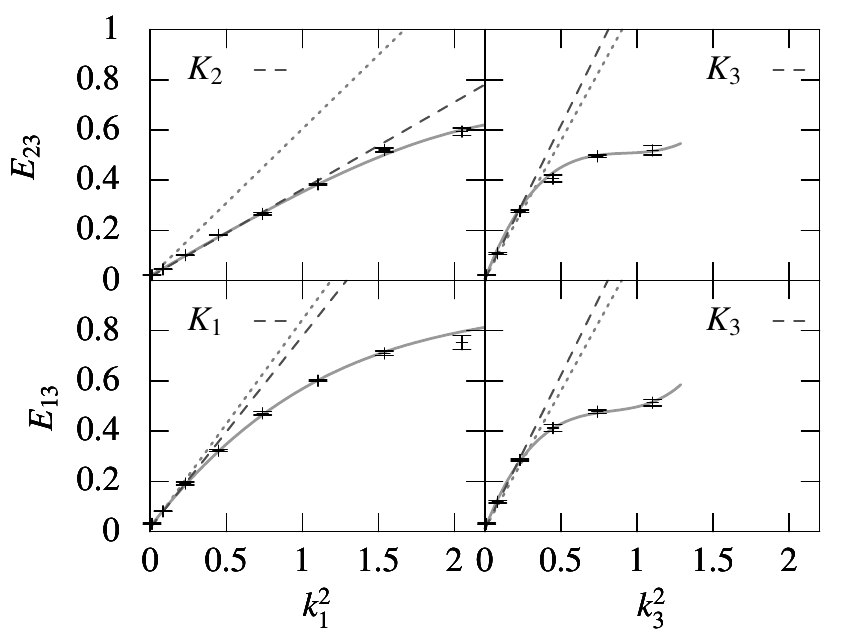}
        \caption{Cuts trough the data along $k_1^2$ (left) and $k_3^2$ (right) at the lowest perpendicular $k^2$-values accounted for in the simulation, i.e., left: $k_3^2=0.01$ and right: $k_1^2=0.01$. 
                 Solid lines indicate the polynomial fit, dashed lines originate from the planar fit, their slopes corresponding to the respective $K_i$ values. Dotted lines show the slopes 
		derived by Allen and Frenkel~\cite{Allen_Calculation:1988,Allen_Erratum:1990} for comparison.}
    \label{fig:Ecuts}
\end{figure}

For the calculation of the Frank elastic constants we have to take into account the fact that our system is enclosed in a box of side length $l$. This allows us to consider only $\vec{k}$ which have 
components that are an integer multiple of the minimal value commensurate with the box in the fixed $xyz$-system, i.e., $\vec{k}=(\kappa_x,\kappa_y,\kappa_z) 2\pi/l$, $\kappa_i\in\mathbb{Z}$. 
As relation \ref{eqn:Emu} only holds for $k\rightarrow0$, it is vital to simulate a box of large enough dimension $l$ to get access to $k$ values small enough to enter the regime 
where $E_{13}(k_1^2,k_3^2)$ and $E_{23}(k_1^2,k_3^2)$ behave linearly; the box size chosen in this simulation proved to be sufficiently large (see Fig.~\ref{fig:Ecuts}).

The Fourier transformed order tensor $\tilde{\boldsymbol{Q}}$ (Eq.~\ref{eqn:FourierOrdertens}) was evaluated in the $xyz$-system every $10000$ time steps for all $\vec{k}$ with $-10<\kappa_i<10$. This provides access to a range of $k^2$ 
large enough to successfully fit the $E_{\mu3}$ data.

\begin{figure}[h!]
        \centering
        \includegraphics[width=8cm]{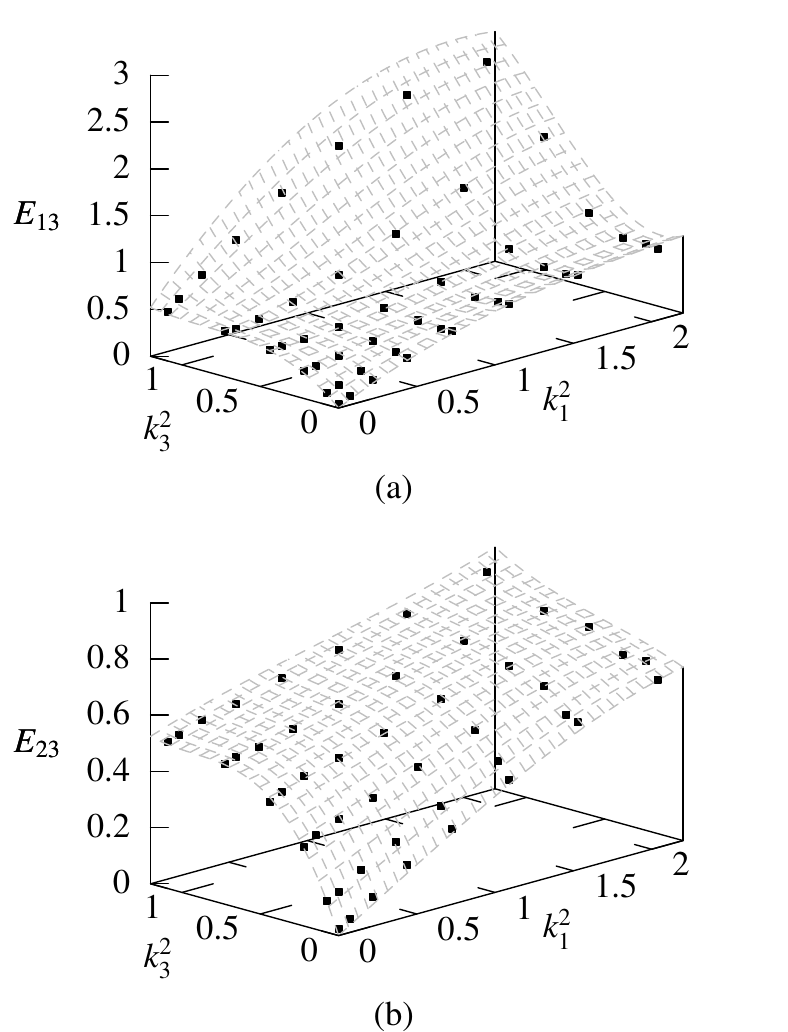}
        \caption{$E_{13}$ (a) and $E_{23}$ (b) as functions of $k_1^2$ and $k_3^2$ (black squares). The respective error bars in $E$ are smaller than the symbol size. 
                The surfaces represent the polynomial fitting functions.}
    \label{fig:Emu3}
\end{figure}

The tensor $\tilde{\boldsymbol{Q}}$ was subsequently transformed to the $123$-system given by the nematic director, calculated as described in Section~\ref{subsec:resOrderparam}, and the currently evaluated $\vec{k}$.
The values for $|\tilde{Q}_{1 3}(\vec{k})|^2$ and $|\tilde{Q}_{2 3}(\vec{k})|^2$ were then averaged over bins of equal size in $k_1$ and $k_3$. 
The results for $E_{\mu3}$ were obtained separately for the different initial configurations from $\left<|\tilde{Q}_{\mu 3}(\vec{k})|^2\right>$ following Eq.~\ref{eqn:Emu}. 
The averaged result of the different runs is shown in Fig.~\ref{fig:Emu3}.

To get a first evaluation of the elastic constants, i.e., the slope of the functions $E_{\mu3}$ at $k^2=0$, the plane $K_\mu k_1^2+K_3 k_3^2$ was fitted to the respective $E_{\mu3}$ data 
in the range of $k_i^2<0.1$. 
Cuts through the data and the corresponding fits are presented in Fig.~\ref{fig:Ecuts}.

To gain a more accurate estimate of the Frank elastic constants, we subsequently fitted the data with bivariate polynomials in $k_1^2$ and $k_3^2$ of the form $\sum_{i,j=0}^3 a_{ij} x^i y^j - a_{00}$ and 
evaluated their leading coefficients. The $K_3$ value gained that way is larger than the one found in the linear 
fit (see Tab.~\ref{tab:Frank}). This indicates that the assumption of a purely linear behavior is not justified even in the range of $k_i^2\leq0.1$.
As the functions are concave in the low $k$ regime, a linear fit will underestimate the derivative at $k^2=0$. From Fig.~\ref{fig:Ecuts} it is clear that the curvature along the $k_3^2$-axis sets in at 
far lower $k^2$ values than along the $k_1^2$-axis. This explains why the values we found for $K_1$ and $K_2$ agree with those derived from the linear fit within the error bars  
while $K_3$ exceeds the one from the linear fit by nearly $30\%$.
\begin{table*}
\begin{threeparttable}[h!]
 \begin{tabular*}{\textwidth}{lcccccc}
\hline
  &$\rho^*$&$P^*$&S&$K_1^*$&$K_2^*$&$K_3^*$\\
\hline\hline\\
  \underline{Present work}&&&&&\\
  $\cdot$ planar fit for $K_i^*$ &\multirow{2}{*}{$0.5$} &\multirow{2}{*}{$6.015\pm0.020$}&\multirow{2}{*}{$0.727\pm0.002$} &$0.763\pm 0.045$&$0.348\pm0.024 $&$1.226\pm0.045$\\
  $\cdot$ polynomial fit for $K_i^*$& & & &$0.812\pm0.019$&$0.352\pm0.016$ &$1.583\pm0.027$\vspace{1ex}\\
  \hline\\
  Frenkel~\cite{Frenkel_Structure:1988} \small{(MC\&MD)}&$0.5$&$5.942$&&&&\vspace{1ex}\\
  Bolhuis~\cite{Bolhuis_Tracing:1997} \small{(MC\&MD)}&$0.5$&$5.995\pm0.059$\tnote{a}&&&&\\
                                                         &$0.513$&$6.4$&$0.74$&&&\vspace{1ex}\\
  McGrother~\cite{McGrother_re‐examination:1996} \small{(MC)}&$0.497\pm0.02$&$5.94$&$0.728\pm0.017$&&&\\
                                           &$0.507\pm0.02$&$6.20$&$0.754\pm0.020$&&&\vspace{4ex}\\
  Allen~\cite{Allen_Calculation:1988}\tnote{b} \small{(MD)}&$0.5$&&$0.73$&$0.83\pm0.25$&$0.59\pm0.07$&$1.10\pm0.11$ 
\vspace{1ex}\\
  Poniewierski~\cite{Poniewierski_Density:1990} \small{(DFT)}&$0.5$&&$0.728$&$0.513$&$0.239$&$1.526$\vspace{1ex}\\
  Somoza~\cite{Somoza_Frank:1989} \small{(DFT)}&$0.5$&&$0.791$&$0.630$&$0.297$&$2.403$\vspace{1ex}\\
  Lee~\cite{Sin-Doo_Density:1989} \small{(DFT)}&$0.482$&&$0.729$&$1.322$&$0.441$&$6.447$ \vspace{1ex}\\ 
 \hline
 \end{tabular*}
 \begin{tablenotes}\small 
\item[a] Data for $P^*$ estimated from Fig.4 Ref.~\cite{Bolhuis_Tracing:1997}.
\item[b] All values and errors of $K_i^*$ have been multiplied by $9/4$ after Ref.~\cite{Allen_Erratum:1990}.
\end{tablenotes}
\caption{Pressure $P^*$ in units of $k_B T/v_0$, Frank elastic constants $K_i^*$ in units of $k_B T/D$ and nematic order parameter $S$ of the hard-spherocylinder system with aspect ratio $L/D=5$ 
at rescaled densities $\rho^*$. Our results are compared to previous data from Monte Carlo~(MC) and molecular dynamics~(MD) simulations as well as from density functional theory (DFT).\\}
\label{tab:Frank}
\end{threeparttable}
\end{table*}\\
The values found in literature for the Frank elastic constants of this system are few and differ considerably. A summary of both MD~\cite{Allen_Calculation:1988,Allen_Erratum:1990} 
and DFT~\cite{Poniewierski_Density:1990,Somoza_Frank:1989,Sin-Doo_Density:1989} results as well as a comparison to our findings is given in Tab.~\ref{tab:Frank}.
As we have chosen the temperature and the length scale of the simulation accordingly, the rescaled results for the
Frank elastic constants $K_i^* =\beta D K_i$ correspond directly to the ones measured in the simulation.

It is clear that our results are well within the range spanned by the data from previous works, showing the expected relation $K_3>K_1>K_2$ found in 
experiments~\cite{Jeu_Determination:1976,Madhusudana_Elasticity:1982} and predicted theoretically~\cite{Priest_Theory:1973}. 
We reason the difference between our results and the simulations by Allen and Frenkel~\cite{Allen_Calculation:1988,Allen_Erratum:1990} to be mainly due to their smaller system size. The minimum $k_i$ 
they could evaluate were considerably larger than those we had access to. Also they could only take the two smallest of their $k_3$ values into account during the fitting procedure as for higher 
values a peak in the structure factor developed. Therefore they used a fitting function that was a product of a second-order polynomial in $k_1$ and a linear function in $k_3$. If we try to fit this function 
to our larger data set, we find that the fit converges very poorly, no matter the $\vec{k}$ range we restrict it to. An additional factor that might contribute to the discrepancy between the results 
is the fact that Allen and Frenkel did not correct for fluctuations of the director, although they judged those to be negligible.

\section{Conclusions} 
\label{sec:conc}
We have shown that the \pe{} physics engine is able to accurately treat the liquid crystal model system comprised of hard spherocylinders with aspect ration $L/D=5$. 
Our results verify the value for the nematic order parameter at a density of $\rho*=0.5$ found in earlier simulations as well as in a number of density functional theory calculations.
In addition we have calculated the pressure in the system which confirms the values published for other simulation techniques. This clearly indicates that already systems with a small 
number of particles compared to the one studied in this work have negligible finite size effects on these two parameters.
 
Finally we have derived the Frank elastic constants at our chosen system parameters. Here the comparison to existing work is more difficult since there is only one set of simulation results available 
and the density functional theory values scatter in a broad range. However, our values certainly provide a better estimate for the elastic constants than the listed simulation data by Allen and Frenkel 
as their comparatively limited system size restricted their access to the low $k$-value regime considerably.

\section*{Acknowledgements}
The work  has been partially funded by the Deutsche Forschungsgemeinschaft (DFG) through the Cluster of Excellence 'Engineering of Advanced Materials' in Erlangen.


\begin{thebibliography}{41}
\expandafter\ifx\csname natexlab\endcsname\relax\def\natexlab#1{#1}\fi
\providecommand{\url}[1]{\texttt{#1}}
\providecommand{\href}[2]{#2}
\providecommand{\path}[1]{#1}
\providecommand{\DOIprefix}{doi:}
\providecommand{\ArXivprefix}{arXiv:}
\providecommand{\URLprefix}{URL: }
\providecommand{\Pubmedprefix}{pmid:}
\providecommand{\doi}[1]{\href{http://dx.doi.org/#1}{\path{#1}}}
\providecommand{\Pubmed}[1]{\href{pmid:#1}{\path{#1}}}
\providecommand{\bibinfo}[2]{#2}
\ifx\xfnm\relax \def\xfnm[#1]{\unskip,\space#1}\fi
\bibitem[{Iglberger and R{\"{u}}de(2009)}]{Iglberger_Rigid:2009}
\bibinfo{author}{K.~Iglberger}, \bibinfo{author}{U.~R{\"{u}}de},
\newblock \bibinfo{title}{Massively parallel rigid body dynamics simulations},
\newblock \bibinfo{journal}{CSRD} \bibinfo{volume}{23} (\bibinfo{year}{2009})
  \bibinfo{pages}{159--167}.
\bibitem[{Iglberger(2010)}]{Iglberger_Software:2010}
\bibinfo{author}{K.~Iglberger}, \bibinfo{title}{{Software Design of a Massively
  Parallel Rigid Body Framework}}, Ph.D. thesis, {Technische Fakult{\"{a}}t der
  Universit{\"{a}}t Erlangen-N{\"{u}}rnberg}, \bibinfo{year}{{2010}}.
\bibitem[{Cundall and Strack(1979)}]{Cundall_dicsrete:1979}
\bibinfo{author}{P.~A. Cundall}, \bibinfo{author}{O.~D.~L. Strack},
\newblock \bibinfo{title}{{A discrete numerical model for granular
  assemblies}},
\newblock \bibinfo{journal}{{G{\'{e}}otechnique}} \bibinfo{volume}{{29}}
  (\bibinfo{year}{{1979}}) \bibinfo{pages}{47--65}.
\bibitem[{Kaufman et~al.(2005)Kaufman, Edmunds, and
  Pai}]{Kaufman_Fast_article:2005}
\bibinfo{author}{D.~M. Kaufman}, \bibinfo{author}{T.~Edmunds},
  \bibinfo{author}{D.~K. Pai},
\newblock \bibinfo{title}{Fast frictional dynamics for rigid bodies},
\newblock \bibinfo{journal}{ACM Trans. Graph.} \bibinfo{volume}{24}
  (\bibinfo{year}{2005}) \bibinfo{pages}{946--956}.
\bibitem[{Jean(1999)}]{Jean:1999}
\bibinfo{author}{M.~Jean},
\newblock \bibinfo{title}{The non-smooth contact dynamics method},
\newblock \bibinfo{journal}{Comput. Method. Appl. M.} \bibinfo{volume}{177}
  (\bibinfo{year}{1999}) \bibinfo{pages}{235--257}.
\bibitem[{Iglberger and R{\"{u}}de(2010)}]{Iglberger_Massively:2010}
\bibinfo{author}{K.~Iglberger}, \bibinfo{author}{U.~R{\"{u}}de},
\newblock \bibinfo{title}{Massively parallel granular flow simulations with
  non-spherical particles},
\newblock \bibinfo{journal}{CSRD} \bibinfo{volume}{25} (\bibinfo{year}{2010})
  \bibinfo{pages}{105--113}.
\bibitem[{Feichtinger et~al.(2011)Feichtinger, Donath, K{\"{o}}stler,
  G{\"{o}}tz, and R{\"{u}}de}]{Feichtinger_walberla:2011}
\bibinfo{author}{C.~Feichtinger}, \bibinfo{author}{S.~Donath},
  \bibinfo{author}{H.~K{\"{o}}stler}, \bibinfo{author}{J.~G{\"{o}}tz},
  \bibinfo{author}{U.~R{\"{u}}de},
\newblock \bibinfo{title}{Walberla: Hpc software design for computational
  engineering simulations},
\newblock \bibinfo{journal}{JOCS} \bibinfo{volume}{2} (\bibinfo{year}{2011})
  \bibinfo{pages}{105--112}.
\bibitem[{G{\"{o}}tz et~al.(2010)G{\"{o}}tz, Iglberger, St{\"{u}}rmer, and
  R{\"{u}}de}]{Goetz_Direct:2010}
\bibinfo{author}{J.~G{\"{o}}tz}, \bibinfo{author}{K.~Iglberger},
  \bibinfo{author}{M.~St{\"{u}}rmer}, \bibinfo{author}{U.~R{\"{u}}de},
\newblock \bibinfo{title}{{Direct Numerical Simulation of Particulate Flows on
  294912 Processor Cores}},
\newblock in: \bibinfo{booktitle}{{High Performance Computing, Networking,
  Storage and Analysis (SC), 2010 International Conference for}},
  \bibinfo{year}{{2010}}, pp. \bibinfo{pages}{1--11}.
  \DOIprefix\doi{{10.1109/SC.2010.20}}.
\bibitem[{Pickl et~al.(2012)Pickl, G{\"{o}}tz, Iglberger, Pande, Mecke, Smith,
  and R{\"{u}}de}]{Pickl_All:2011}
\bibinfo{author}{K.~Pickl}, \bibinfo{author}{J.~G{\"{o}}tz},
  \bibinfo{author}{K.~Iglberger}, \bibinfo{author}{J.~Pande},
  \bibinfo{author}{K.~Mecke}, \bibinfo{author}{A.-S. Smith},
  \bibinfo{author}{U.~R{\"{u}}de},
\newblock \bibinfo{title}{{All good things come in threes{---}Three beads learn
  to swim with lattice Boltzmann and a rigid body solver }},
\newblock \bibinfo{journal}{{JOCS}} \bibinfo{volume}{{3}}
  (\bibinfo{year}{{2012}}) \bibinfo{pages}{374--387}.
\bibitem[{Preclik(tted)}]{Preclik_Ultrascale:2014}
\bibinfo{author}{T.~Preclik}, \bibinfo{title}{{Models and Algorithms for
  Ultrascale Simulations of Non-smooth Granular Dynamics}}, Ph.D. thesis,
  {Technische Fakult{\"{a}}t der Universit{\"{a}}t Erlangen-N{\"{u}}rnberg},
  \bibinfo{year}{submitted}.
\bibitem[{Moreau and Panagiotopoulos(1988)}]{Moreau_Panagiotopoulos:1988}
\bibinfo{author}{J.~Moreau}, \bibinfo{author}{P.~Panagiotopoulos},
  \bibinfo{title}{Nonsmooth Mechanics and Applications}, volume
  \bibinfo{volume}{302}, \bibinfo{publisher}{Springer}, \bibinfo{year}{1988}.
\bibitem[{Few and Rigby(1973)}]{Few_Equation:1973}
\bibinfo{author}{G.~Few}, \bibinfo{author}{M.~Rigby},
\newblock \bibinfo{title}{Equation of state for systems of hard non-spherical
  molecules},
\newblock \bibinfo{journal}{Chem. Phys. Lett.} \bibinfo{volume}{20}
  (\bibinfo{year}{1973}) \bibinfo{pages}{433 -- 435}.
\bibitem[{Vieillard-Baron(1974)}]{Vieillard-Baron_equation:1974}
\bibinfo{author}{J.~Vieillard-Baron},
\newblock \bibinfo{title}{{The equation of state of a system of hard
  spherocylinders}},
\newblock \bibinfo{journal}{{Mol. Phys.}} \bibinfo{volume}{{28}}
  (\bibinfo{year}{{1974}}) \bibinfo{pages}{809--818}.
\bibitem[{Bernal and Fankuchen(1941)}]{Bernal_xray:1941}
\bibinfo{author}{J.~D. Bernal}, \bibinfo{author}{I.~Fankuchen},
\newblock \bibinfo{title}{{X-RAY AND CRYSTALLOGRAPHIC STUDIES OF PLANT VIRUS
  PREPARATIONS I. INTRODUCTION AND PREPARATION OF SPECIMENS II. MODES OF
  AGGREGATION OF THE VIRUS PARTICLES}},
\newblock \bibinfo{journal}{J. Gen. Physiol.} \bibinfo{volume}{25}
  (\bibinfo{year}{1941}) \bibinfo{pages}{111--146}.
\bibitem[{Kuijk et~al.(2012)Kuijk, Byelov, Petukhov, van Blaaderen, and
  Imhof}]{Kuijk_phase:2012}
\bibinfo{author}{A.~Kuijk}, \bibinfo{author}{D.~V. Byelov},
  \bibinfo{author}{A.~V. Petukhov}, \bibinfo{author}{A.~van Blaaderen},
  \bibinfo{author}{A.~Imhof},
\newblock \bibinfo{title}{Phase behavior of colloidal silica rods},
\newblock \bibinfo{journal}{Faraday Discuss.} \bibinfo{volume}{159}
  (\bibinfo{year}{2012}) \bibinfo{pages}{181--199}.
\bibitem[{Onsager(1949)}]{Onsager_effects:1949}
\bibinfo{author}{L.~Onsager},
\newblock \bibinfo{title}{The effects of shape on the interaction of colloidal
  particles},
\newblock \bibinfo{journal}{Ann. NY Acad. Sci.} \bibinfo{volume}{51}
  (\bibinfo{year}{1949}) \bibinfo{pages}{627--659}.
\bibitem[{Frenkel(1988)}]{Frenkel_Structure:1988}
\bibinfo{author}{D.~Frenkel},
\newblock \bibinfo{title}{Structure of hard-core models for liquid crystals},
\newblock \bibinfo{journal}{J. Phys. Chem.} \bibinfo{volume}{92}
  (\bibinfo{year}{1988}) \bibinfo{pages}{3280--3284}.
\bibitem[{Frenkel et~al.(1988)Frenkel, Lekkerkerker, and
  Stroobants}]{Frenkel_Thermodynamic:1988}
\bibinfo{author}{D.~Frenkel}, \bibinfo{author}{H.~N.~W. Lekkerkerker},
  \bibinfo{author}{A.~Stroobants},
\newblock \bibinfo{title}{Thermodynamic stability of a smectic phase in a
  system of hard rods},
\newblock \bibinfo{journal}{Nature} \bibinfo{volume}{332}
  (\bibinfo{year}{1988}) \bibinfo{pages}{822--823}.
\bibitem[{Veerman and Frenkel(1990)}]{Veerman_Phase:1990}
\bibinfo{author}{J.~A.~C. Veerman}, \bibinfo{author}{D.~Frenkel},
\newblock \bibinfo{title}{Phase diagram of a system of hard spherocylinders by
  computer simulation},
\newblock \bibinfo{journal}{Phys. Rev. A} \bibinfo{volume}{41}
  (\bibinfo{year}{1990}) \bibinfo{pages}{3237--3244}.
\bibitem[{Bolhuis and Frenkel(1997)}]{Bolhuis_Tracing:1997}
\bibinfo{author}{P.~Bolhuis}, \bibinfo{author}{D.~Frenkel},
\newblock \bibinfo{title}{Tracing the phase boundaries of hard
  spherocylinders},
\newblock \bibinfo{journal}{J. Chem. Phys.} \bibinfo{volume}{106}
  (\bibinfo{year}{1997}) \bibinfo{pages}{666--687}.
\bibitem[{McGrother et~al.(1996)McGrother, Williamson, and
  Jackson}]{McGrother_re‐examination:1996}
\bibinfo{author}{S.~C. McGrother}, \bibinfo{author}{D.~C. Williamson},
  \bibinfo{author}{G.~Jackson},
\newblock \bibinfo{title}{A re‐examination of the phase diagram of hard
  spherocylinders},
\newblock \bibinfo{journal}{J. Chem. Phys.} \bibinfo{volume}{104}
  (\bibinfo{year}{1996}) \bibinfo{pages}{6755--6771}.
\bibitem[{Kaufman et~al.(2005)Kaufman, Edmunds, and
  Pai}]{Kaufman_Fast_proceedings:2005}
\bibinfo{author}{D.~M. Kaufman}, \bibinfo{author}{T.~Edmunds},
  \bibinfo{author}{D.~K. Pai},
\newblock \bibinfo{title}{Fast frictional dynamics for rigid bodies},
\newblock in: \bibinfo{booktitle}{ACM SIGGRAPH 2005 Papers}, SIGGRAPH '05,
  \bibinfo{publisher}{ACM}, \bibinfo{address}{New York, NY, USA},
  \bibinfo{year}{2005}, pp. \bibinfo{pages}{946--956}. \URLprefix
  \url{http://doi.acm.org/10.1145/1186822.1073295}.
  \DOIprefix\doi{10.1145/1186822.1073295}.
\bibitem[{Gropp et~al.(1999)Gropp, Lusk, and Skjellum}]{Gropp_Using:1999}
\bibinfo{author}{W.~Gropp}, \bibinfo{author}{E.~Lusk},
  \bibinfo{author}{A.~Skjellum}, \bibinfo{title}{Using MPI (2nd ed.): portable
  parallel programming with the message-passing interface},
  \bibinfo{publisher}{MIT Press}, \bibinfo{address}{Cambridge, MA, USA},
  \bibinfo{year}{1999}.
\bibitem[{McNamara and Young(1994)}]{McNamara_Inelastic:1994}
\bibinfo{author}{S.~McNamara}, \bibinfo{author}{W.~R. Young},
\newblock \bibinfo{title}{Inelastic collapse in two dimensions},
\newblock \bibinfo{journal}{Phys. Rev. E} \bibinfo{volume}{50}
  (\bibinfo{year}{1994}) \bibinfo{pages}{R28--R31}.
\bibitem[{de~Gennes(1974)}]{Gennes:1974}
\bibinfo{author}{P.~G. de~Gennes}, \bibinfo{title}{The Physics of Liquid
  Crystals}, \bibinfo{publisher}{Clarendon Press Oxford}, \bibinfo{year}{1974}.
\bibitem[{Eppenga and Frenkel(1984)}]{Eppenga:1984}
\bibinfo{author}{R.~Eppenga}, \bibinfo{author}{D.~Frenkel},
\newblock \bibinfo{title}{Monte carlo study of the isotropic and nematic phases
  of infinitely thin hard platelets},
\newblock \bibinfo{journal}{Mol. Phys.} \bibinfo{volume}{52}
  (\bibinfo{year}{1984}) \bibinfo{pages}{1303--1334}.
\bibitem[{Frenkel and Smit(1996)}]{Frenkel_Understanding:1996}
\bibinfo{author}{D.~Frenkel}, \bibinfo{author}{B.~Smit},
  \bibinfo{title}{Understanding Molecular Simulation},
  \bibinfo{publisher}{Academic Press Elsevier}, \bibinfo{year}{1996}.
\bibitem[{Rebertus and Sando(1977)}]{Rebertus_Molecular:1977}
\bibinfo{author}{D.~W. Rebertus}, \bibinfo{author}{K.~M. Sando},
\newblock \bibinfo{title}{Molecular dynamics simulation of a fluid of hard
  spherocylinders},
\newblock \bibinfo{journal}{J. Chem. Phys.} \bibinfo{volume}{67}
  (\bibinfo{year}{1977}) \bibinfo{pages}{2585}.
\bibitem[{Frank(1958)}]{Frank_Liquid:1958}
\bibinfo{author}{F.~C. Frank},
\newblock \bibinfo{title}{I. liquid crystals. on the theory of liquid
  crystals},
\newblock \bibinfo{journal}{Discuss. Faraday Soc.} \bibinfo{volume}{25}
  (\bibinfo{year}{1958}) \bibinfo{pages}{19--28}.
\bibitem[{Allen and Frenkel(1988)}]{Allen_Calculation:1988}
\bibinfo{author}{M.~P. Allen}, \bibinfo{author}{D.~Frenkel},
\newblock \bibinfo{title}{Calculation of liquid-crystal frank constants by
  computer simulation},
\newblock \bibinfo{journal}{Phys. Rev. A} \bibinfo{volume}{37}
  (\bibinfo{year}{1988}) \bibinfo{pages}{1813}.
\bibitem[{Allen and Frenkel(1990)}]{Allen_Erratum:1990}
\bibinfo{author}{M.~P. Allen}, \bibinfo{author}{D.~Frenkel},
\newblock \bibinfo{title}{Erratum: Calculation of liquid-crystal frank
  constants by computer simulation [phys. rev. a \textbf{37}, 1813 (1988)]},
\newblock \bibinfo{journal}{Phys. Rev. A} \bibinfo{volume}{42}
  (\bibinfo{year}{1990}) \bibinfo{pages}{3641--3641}.
\bibitem[{Tjipto-Margo et~al.(1992)Tjipto-Margo, Evans, Allen, and
  Frenkel}]{TjiptoMargo_Elastic:1992}
\bibinfo{author}{B.~Tjipto-Margo}, \bibinfo{author}{G.~T. Evans},
  \bibinfo{author}{M.~P. Allen}, \bibinfo{author}{D.~Frenkel},
\newblock \bibinfo{title}{{Elastic constants of hard and soft nematic liquid
  crystals}},
\newblock \bibinfo{journal}{{J. Phys. Chem.}} \bibinfo{volume}{{96 }}
  (\bibinfo{year}{{1992}}) \bibinfo{pages}{3942{--}3948}.
\bibitem[{Allen et~al.(1996)Allen, Warren, Wilson, Sauron, and
  Smith}]{Allen_Molecular:1996}
\bibinfo{author}{M.~P. Allen}, \bibinfo{author}{M.~A. Warren},
  \bibinfo{author}{M.~R. Wilson}, \bibinfo{author}{A.~Sauron},
  \bibinfo{author}{W.~Smith},
\newblock \bibinfo{title}{{Molecular dynamics calculation of elastic constants
  in Gay{--}Berne nematic liquid crystals}},
\newblock \bibinfo{journal}{{J. Chem. Phys.}} \bibinfo{volume}{{105}}
  (\bibinfo{year}{{1996}}) \bibinfo{pages}{2850}.
\bibitem[{O'Brien et~al.(2008)O'Brien, Allen, Cheung, Dennison, and
  Masters}]{OBrien_Elastic:2008}
\bibinfo{author}{P.~A. O'Brien}, \bibinfo{author}{M.~P. Allen},
  \bibinfo{author}{D.~L. Cheung}, \bibinfo{author}{M.~Dennison},
  \bibinfo{author}{A.~Masters},
\newblock \bibinfo{title}{{Elastic constants of hard thin platelets by Monte
  Carlo simulation and virial expansion}},
\newblock \bibinfo{journal}{{Phys. Rev. E}} \bibinfo{volume}{{78}}
  (\bibinfo{year}{{2008}}).
\bibitem[{Forster(1975)}]{Forster_Hydrodynamic:1975}
\bibinfo{author}{D.~Forster}, \bibinfo{title}{Hydrodynamic Fluctuations, Broken
  Symmetry and Correlation Functions}, \bibinfo{publisher}{Frontiers in
  Physics}, \bibinfo{year}{1975}.
\bibitem[{Poniewierski and Ho{\l}yst(1990)}]{Poniewierski_Density:1990}
\bibinfo{author}{A.~Poniewierski}, \bibinfo{author}{R.~Ho{\l}yst},
\newblock \bibinfo{title}{{Density-functional theory for systems of hard
  rods}},
\newblock \bibinfo{journal}{{Phys. Rev. A}} \bibinfo{volume}{{41}}
  (\bibinfo{year}{{1990}}) \bibinfo{pages}{6871--6880}.
\bibitem[{Lee(1989)}]{Sin-Doo_Density:1989}
\bibinfo{author}{S.-D. Lee},
\newblock \bibinfo{title}{{Density-functional approach to curvature elasticity
  in a liquid-density nematic system}},
\newblock \bibinfo{journal}{{Phys. Rev. A}} \bibinfo{volume}{{39}}
  (\bibinfo{year}{{1989}}) \bibinfo{pages}{3631--3639}.
\bibitem[{Somoza and Tarazona(1989)}]{Somoza_Frank:1989}
\bibinfo{author}{A.~M. Somoza}, \bibinfo{author}{P.~Tarazona},
\newblock \bibinfo{title}{{Frank elastic constants of a nematic liquid crystal
  of hard molecules}},
\newblock \bibinfo{journal}{{Phys. Rev. A}} \bibinfo{volume}{{40}}
  (\bibinfo{year}{{1989}}) \bibinfo{pages}{6069--6076}.
\bibitem[{De~Jeu et~al.(1976)De~Jeu, Claassen, and
  Spruijt}]{Jeu_Determination:1976}
\bibinfo{author}{W.~H. De~Jeu}, \bibinfo{author}{W.~A.~P. Claassen},
  \bibinfo{author}{A.~M.~J. Spruijt},
\newblock \bibinfo{title}{The determination of the elastic constants of nematic
  liquid crystals},
\newblock \bibinfo{journal}{Mol. Cryst. Liq. Cryst.} \bibinfo{volume}{37}
  (\bibinfo{year}{1976}) \bibinfo{pages}{269--280}.
\bibitem[{Madhusudana and Pratibha(1982)}]{Madhusudana_Elasticity:1982}
\bibinfo{author}{N.~V. Madhusudana}, \bibinfo{author}{R.~Pratibha},
\newblock \bibinfo{title}{Elasticity and orientational order in some
  cyanobiphenyls: Part iv. reanalysis of the data},
\newblock \bibinfo{journal}{Mol. Cryst. Liq. Cryst.} \bibinfo{volume}{89}
  (\bibinfo{year}{1982}) \bibinfo{pages}{249--257}.
\bibitem[{Priest(1973)}]{Priest_Theory:1973}
\bibinfo{author}{R.~G. Priest},
\newblock \bibinfo{title}{Theory of the frank elastic constants of nematic
  liquid crystals},
\newblock \bibinfo{journal}{Phys. Rev. A} \bibinfo{volume}{7}
  (\bibinfo{year}{1973}) \bibinfo{pages}{720--729}.

\end{thebibliography}


\end{document}